\documentclass[aps,prd,twocolumn,showpacs,floatfix,preprintnumbers,amsfont,amsmath,amssymb,nofootinbib,superscriptaddress]{revtex4-1}

\usepackage{aas_macros}
\usepackage{graphicx}
\usepackage{dcolumn}
\usepackage{bm}
\usepackage[normalem]{ulem}
\usepackage[usenames,dvipsnames,svgnames,table]{xcolor}

\usepackage{xcolor}

\definecolor{colorLink}{rgb}{0.9,0,0} 
\definecolor{colorCite}{rgb}{0,0.7,0} 
\definecolor{colorURL} {rgb}{0,0,0.8} 
\usepackage[colorlinks=true,linktocpage=true,linkcolor=colorLink,citecolor=colorCite,urlcolor=colorURL]{hyperref}
\usepackage{bbold}
\usepackage{multirow}
\usepackage{fontawesome}
\usepackage{algorithm2e}
\usepackage{algpseudocode}

\newcommand{\be}{\begin{equation}}
\newcommand{\ee}{\end{equation}}

\newcommand{\sk}[1]{}

\newcommand{\nbicon}{\faFileCodeO}
\newcommand{\nblink}{\href{https://github.com/JayWadekar/GW_higher_harmonics_search/tree/main/Pipeline_modules_arXiv_2405.1740}{\nbicon}}

\def\msun{\rm M_{\odot}}
\def\M{\mathcal{M}}
\def\h{\mathbb{h}}
\DeclareMathSymbol{\mhyphen}{\mathord}{AMSa}{"39}

\graphicspath{{./}{figs/}}

\begin{document}

\title{New search pipeline for gravitational waves with higher-order modes using mode-by-mode filtering}

 \author{Digvijay Wadekar}
  \email{jayw@ias.edu}
 \affiliation{\mbox{School of Natural Sciences, Institute for Advanced Study, 1 Einstein Drive, Princeton, NJ 08540, USA}}
 \author{Tejaswi Venumadhav}
\affiliation{\mbox{Department of Physics, University of California at Santa Barbara, Santa Barbara, CA 93106, USA}}
\affiliation{\mbox{International Centre for Theoretical Sciences, Tata Institute of Fundamental Research, Bangalore 560089, India}}
 \author{Javier Roulet}
\affiliation{TAPIR, Walter Burke Institute for Theoretical Physics, California Institute of Technology, Pasadena, CA 91125, USA}
\author{Ajit Kumar Mehta}
\affiliation{\mbox{Department of Physics, University of California at Santa Barbara, Santa Barbara, CA 93106, USA}}
\author{Barak Zackay}
\affiliation{\mbox{Department of Particle Physics \& Astrophysics, Weizmann Institute of Science, Rehovot 76100, Israel}}
\author{Jonathan Mushkin}
\affiliation{\mbox{Department of Particle Physics \& Astrophysics, Weizmann Institute of Science, Rehovot 76100, Israel}}
  \author{Matias Zaldarriaga}
\affiliation{\mbox{School of Natural Sciences, Institute for Advanced Study, 1 Einstein Drive, Princeton, NJ 08540, USA}}

 \date{May 15, 2024}

\begin{abstract}
Nearly all template--based gravitational wave (GW) searches only include the quasi-circular quadrupolar modes of the signals in their templates.
Including additional degrees of freedom in the GW templates corresponding to higher-order harmonics, orbital precession, or eccentricity is challenging because: ($i$) the size of template banks and the matched-filtering cost increases significantly with the number of degrees of freedom, $(ii)$ if these additional degrees are not included properly, the search can lose sensitivity overall (due to an increase in the rate of background triggers). 
Here, we focus on including aligned-spin higher harmonics in GW search templates. We use a new mode-by-mode filtering approach, where we separately filter GW strain data with three harmonics [namely $(\ell, |m|)=(2,2)$, $(3,3)$ and $(4,4)$]. 
This results in an increase in the matched-filtering cost by only a factor of $3$ compared to that of a $(2,2)$-only search. 
We develop computationally cheap trigger-ranking statistics to optimally combine the different signal-to-noise ratios (SNR) timeseries from different harmonics, which ensure only physically-allowed combinations of the different harmonics are triggered on. 
We use an empirical template-dependent background model in our ranking statistic to account for non-Gaussian transients. 
In addition, we develop a tool called band eraser which specifically excises narrow time-varying noisy bands in time-frequency space (without having to excise entire time chunks in the data).
New GW candidate events that we detect using our \texttt{IAS-HM} search pipeline and the details of our template banks are discussed in accompanying papers: \citet{Wad23_HM_Events} and \cite{Wad23_TemplateBanks} respectively.  Apart from higher harmonics, we expect our methodology to also be useful for cheap and optimal searches including orbital precession and eccentricity in GW waveforms.
\href{https://github.com/JayWadekar/gwIAS-HM}{\faGithub}

\end{abstract}
\maketitle

\section{Introduction}

Most of the gravitational wave mergers reported until now have been detected by searches which use GW template waveforms \cite{GW230529, O1catalog_LVC2016, gwtc1_o2catalog_LVC2018, lvc_o3a_gwtc2_catalog_2021, lvc_o3a_deep_gwtc2_1_update_2021, lvc_gwtc3_o3_ab_catalog_2021, LVK_O3_IMBH_search, gstlal, PYCBCPipeline, mbta_o3a_pastro_andres2022, ias_pipeline_o1_catalog_new_search_prd2019, ias_o2_pipeline_new_events_prd2020, Ols22_ias_o3a, NitzCatalog_1-OGC_o1_2018, NitzCatalog_2-OGC_o2_2020, nitz_o3a_3ogc_catalog_2021, nitz_4ogc_o3_ab_catalog_2021,Chi23, Meh23_ias_o3b, Kum24}. In such searches, first a template bank is constructed containing GW waveforms corresponding to different binary parameters. The templates are then filtered with the data and triggers with relatively high signal-to-noise ratio (SNR) are saved.

There is an important trade-off to consider when designing template-based searches. On one hand, we want to include as many physical effects in the templates as possible. On the other hand, as we allow more degrees of freedom for waveforms in the template bank, the size of the template bank (and the cost of the search) can increase by multiple orders of magnitude. This is the primary reason why nearly all current templated searches only use simple quasi-circular quadrupole mode waveforms and neglect higher-order harmonics, spin-orbit precession or orbital eccentricity (with only a handful of exceptions, e.g., \cite{Cha22, McI23, Sch24_Precession_O3_Search, Nit20_eccentricity_search, Dhu23_eccentricity_search}).

In this paper, we focus on including aligned-spin higher-order modes (HM) in an efficient way in the templates. Unlike $(2,2)$-only searches, one needs to take into account variation in inclination and the initial orbital phase ($\iota$, $\phi_\mathrm{initial}$) while searching for HM as changing these parameters leads to a relative change in the amplitude or phases of higher harmonics compared to that of the $(2,2)$ mode.
A brute-force approach would involve sampling over both intrinsic $(m_1,m_2, \chi_{1z},\chi_{2z})$ and extrinsic parameters ($\iota$, $\phi_\mathrm{initial}$) to create template bank waveforms. However, this can lead to size of the template bank to increase by $\sim 100 \times$ compared to the $(2,2)$-only case \cite{Cha22, Sch23_NF_TemplateBank, Har18}.

There is another potential problem with introducing new degrees of freedom in the templates.
If these new degrees of freedom are not properly treated, it will lead to loss of sensitivity of the search \cite{Cap14, McI23, Sch24_Precession_Pipeline}. For example, if we sample over the binary inclination to create template bank waveforms, we would need to add smaller weights to the edge-on configurations instead of face-on configuration (due to a difference in their observable volume), otherwise sensitivity of the search can be reduced.

\begin{figure*}
\centering
\includegraphics[width=\textwidth,keepaspectratio=true]{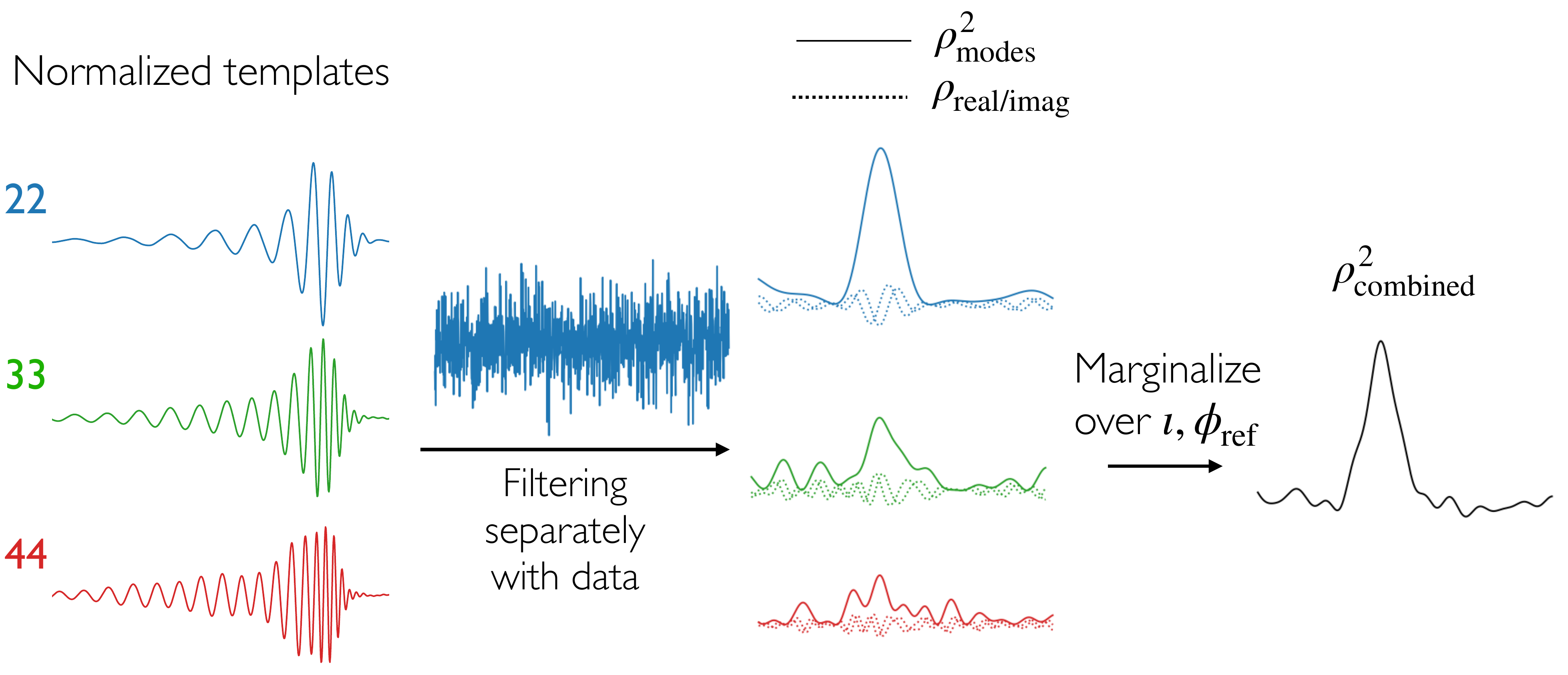}
\caption{ We exploit the fact that the inner product of the data with the full waveform can be written down as a linear combination of the inner product with the separate harmonics.
In our template banks, for each set of intrinsic parameters (i.e., masses and spins), we generate and store the normalized waveforms ($\h_{22}, \h_{33}, \h_{44}$) separately (see section~\ref{sec:TemplateBanks}). We then filter the data with the individual modes, which results in a cost increase of just $3\times$ compared to the (2,2)-only case (instead of 100$\times$ cost incurred when templates are made by combining different harmonics \cite{Cha22, Sch23_NF_TemplateBank, Har18}). We store the output SNR timeseries of each harmonic ($\rho_{\ell m} = \langle \h_{\ell m}| \mathrm{data} \rangle$). In section~\ref{sec:CoherentScore} and Appendix~\ref{sec:single_det_statistic}, we discuss methods to combine the three SNR timeseries by marginalizing over inclination and the initial phase of the binary ($\iota, \phi_0$). This figure was originally presented in our companion paper for HM template banks \cite{Wad23_TemplateBanks} and is provided here for completeness.}
\label{fig:Triggering_modes}
\end{figure*}

We overcome the above issues by using a new mode-by-mode filtering approach to search with HM in this paper and our companion papers \cite{Wad23_HM_Events} and \cite{Wad23_TemplateBanks} (see Fig.~\ref{fig:Triggering_modes}). We first generate normalized templates for three different harmonics [namely $(2,2)$, $(3,3)$ and $(4,4)$] by sampling only over intrinsic binary parameters \cite{Wad23_TemplateBanks}. We then filter the GW strain data separately with these harmonics and collect the three SNR timeseries. This results in an increase in the matched-filtering cost by only a factor of $3$ with respect to that of a $(2,2)$-only search.

In this paper, we develop computationally-cheap detection statistics for optimally combining the different signal-to-noise ratios (SNR) timeseries from each harmonic. We only allow physical combinations of the different harmonics (corresponding to physical configurations of $\iota$ and $\phi_\mathrm{initial}$ and intrinsic binary parameters) and appropriately penalize the unphysical combinations. Thus, apart from a reduction in the matched-filtering cost in our method, the methodology that we develop here also helps us prevent a loss in sensitivity. It is worth noting that our statistics enable optimal marginalization over $\iota$ and $\phi_\mathrm{initial}$, which is difficult to do if $\iota$ and $\phi_\mathrm{initial}$ are explicitly encoded in the templates.

Other than improving the IAS pipeline by adding HM, we also develop new methods to downweight/remove triggers originating from non-Gaussian noise. This becomes especially important when the duration of the waveform in detector band is small (for high-mass binaries which merge at low frequencies) and short-duration instrumental transients (glitches) can mimic real signals \cite{Sha22}. This is also the region where HM have a large contribution to SNR as the bulk of the $(2,2)$ signal is downweighted due to increase in the noise power spectral density at low frequencies \cite{Wad23_TemplateBanks, HMeffect_ParameterSpaceDependency_PekowskyPRD2013, HMeffect_RelativeModeSignificance_HealyPRD2013,  Cap14, HMeffect_AlignedSearchImpactCalderonBustilloPRD2016, HMandPrecessionEffect_HeavySearchImpact_CalderonBustilloPRD2017, HMeffect_IMBHsearchImpact_CalderonBustilloPRD2018, Mil21, Har18, Cha22, Sha22,Zha23}.
We ran the \texttt{IAS-HM} pipeline described here on the public LIGO--Virgo data from the third observing run (O3). In a companion paper \cite{Wad23_HM_Events}, we report new candidate events that we discover as well as previously known events which we recover. In particular, we find new candidate events at high masses and high redshifts as the relative contribution of HM to SNR becomes larger for such systems.

We first provide a brief outline of our template banks in section~\ref{sec:TemplateBanks}. We then derive our trigger-ranking statistics in section~\ref{sec:NewDetStats}, where we first derive our marginalized statistic under the assumption of Gaussian noise in section~\ref{sec:CoherentScore} (for collecting triggers before ranking them, we use a different statistic which is cheaper to compute and is derived in Appendix~\ref{sec:single_det_statistic}). We then derive the correction factor in our ranking statistics due to non-Gaussian background in section~\ref{sec:nonGaussian_statistic}. We discuss improvements to our data pre-processing techniques and introduce a new tool called band eraser in section~\ref{sec:DataPreproc}. We discuss compatibility of our techniques with other searches in the literature and also dependence on assumed astrophysical prior in section~\ref{sec:Discussion} and we conclude in section~\ref{sec:Conclusions}.



\begin{figure*}
\centering
\includegraphics[scale=0.7,keepaspectratio=true]{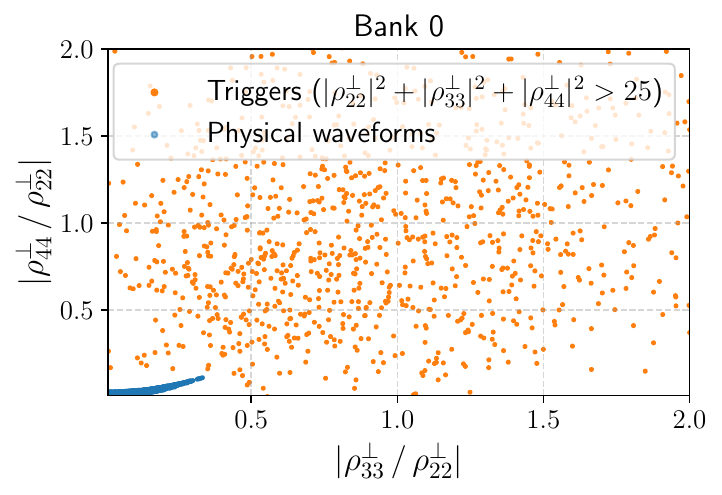}
\includegraphics[scale=0.7,keepaspectratio=true]{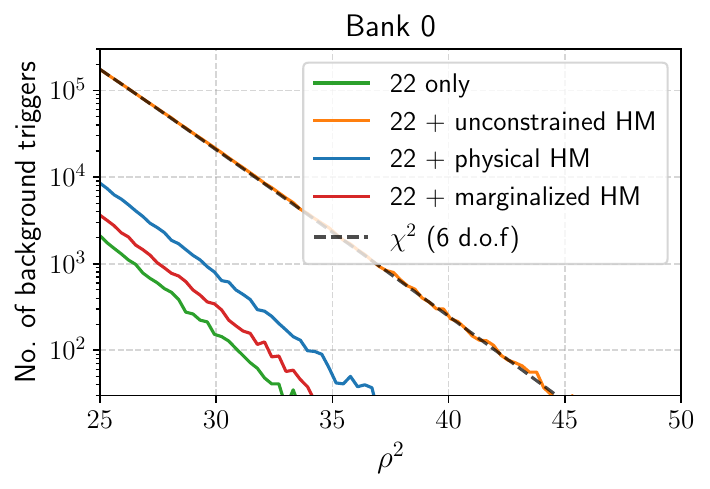}
\includegraphics[scale=0.7,keepaspectratio=true]{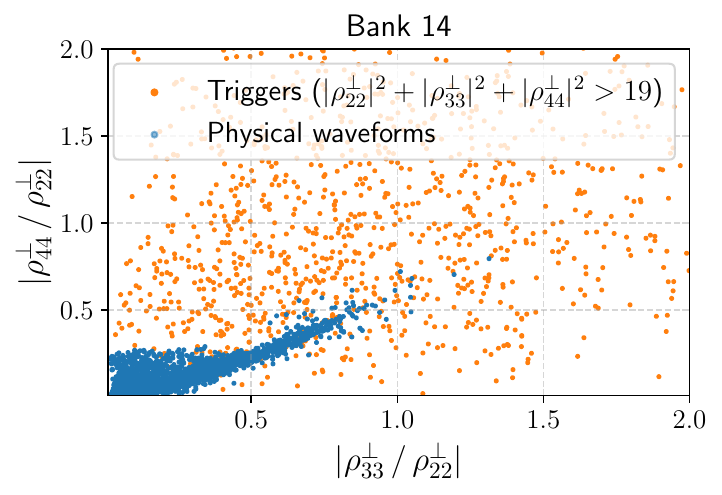}
\includegraphics[scale=0.7,keepaspectratio=true]{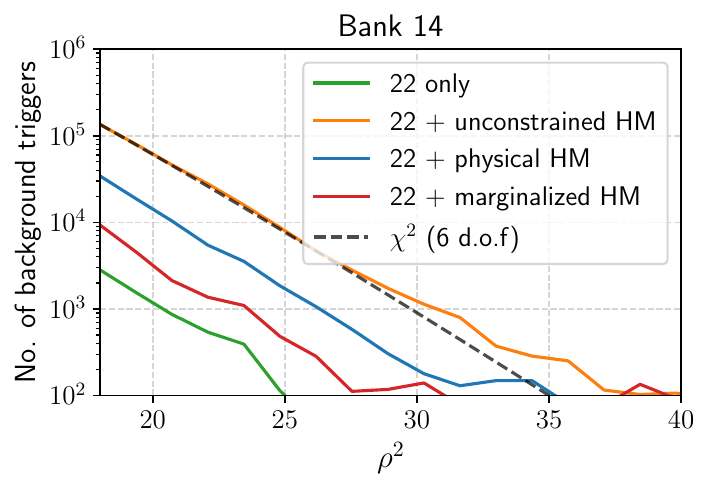}
\caption{\textbf{Top left:} We matched-filter the strain in a particular O3a file with normalized and orthogonalized templates corresponding to different modes and show SNR ratios of the resulting triggers as orange points. Most of these triggers are expected to be due to Gaussian noise (which can have arbitrarily large values of $|\rho^\perp_{33}/\rho^\perp_{22}|$ and $|\rho^\perp_{44}/\rho^\perp_{22}|$). We also show in blue points the SNR ratios for physical injections within the parameter range of our lowest mass bank.
As the parameter space covered by the two sets is vastly different, we devise trigger collection and ranking statistics in Appendix~\ref{sec:single_det_statistic} and section~\ref{sec:CoherentScore} which penalize triggers based on their distance from the blue points.\\
\textbf{Top right:} 
Number of triggers collected using different statistics. Collecting the triggers naively using $\rho^2\equiv|\rho^\perp_{22}|^2+|\rho^\perp_{33}|^2+|\rho^\perp_{44}|^2$ (orange) will allow many more background triggers as compared to only using $\rho^2\equiv|\rho_{22}|^2$ (green).
If we penalize a trigger based only on the distance from the nearest physical injection, we get the blue curve. However, in the red curve, we show the result from an optimal statistic, where we effectively marginalize over the overlap with all the physical injections shown in the left panel. See Eq.~\eqref{eq:SingleDetScore} for the single-detector marginalized-score that we use to collect triggers (which is an approximate version of the full ranking statistic derived in section~\ref{sec:CoherentScore}).
Using the marginalized ranking-statistic not only helps in preserving the sensitivity of our HM pipeline and also decreases our overall search cost (as the number of triggers collected and processed is significantly reduced).\\
\textbf{Bottom panels:} Same as the top panels except for a high mass bank, where relatively higher values of mode SNR ratios are physically possible. \nblink
}
\label{fig:Mode_ratios}
\end{figure*}



\section{Template banks}
\label{sec:TemplateBanks}

We discuss in detail our Fourier-space HM bank construction in a companion paper, \citet{Wad23_TemplateBanks}.
Here, we briefly summarize the important details and reiterate the formalism which we will use later in section~\ref{sec:NewDetStats} for deriving our trigger-ranking statistics. 
To calculate the overlap of two waveforms when we construct the template bank, we use
\be
\langle h_i | h_j \rangle = 4 \int_0^\infty {\rm d}f \frac{h^*_i(f)h_j(f)}{S_n(f)}
\label{eq:inner_prod}
\ee
where $S_n(f)$ is the one-sided power spectral density (PSD).
We search over the following parameter space
\be\begin{split}
3 \,\msun < m^\mathrm{det}_2 &< m^\mathrm{det}_1 < 400\, \msun \\
1/18 &< q < 1\\
|\chi_1|, |\chi_2| &<0.99
\end{split}\label{eq:prior}\ee
and we divide it into 17 banks based on the normalized amplitude shape of the $(2,2)$ waveform (see Fig.~3 of \cite{Wad23_TemplateBanks}). Our $(2,2)$ template bank construction is based on the geometric placement methodology of \cite{ias_template_bank_PSD_roulet2019}, but with a notable improvement that we compress the dimensionality of our banks using a machine learning method (as we will describe later in this section).

Our template bank construction is motivated by the stationary phase approximation (SPA) based results from Ref.~\cite{Mis16}.
At the lowest order, the PN expansion for the frequency-domain amplitudes of HM gives \cite{Wad23_TemplateBanks}
\be\begin{split}
\left|\frac{h_{33}(3f)}{h_{22}(2f)}\right| \simeq& \frac{3\sqrt{5}}{2\sqrt{7}}\, \left[\frac{1-q}{1+q}\right]\, (2\pi M_\mathrm{tot} f)^{1/3}\, \sin (\iota)\\
\left|\frac{h_{44}(4f)}{h_{22}(2f)}\right| \simeq& \frac{8\sqrt{10}}{9\sqrt{7}}\, \left[1-\frac{3 q}{(1+q)^2}\right](2\pi M_\mathrm{tot} f)^{2/3}\, \sin^2 (\iota)
\end{split}
\label{eq:AmpFormulae}\ee
where $q$ is the mass ratio ($m_2/m_1$) and $\iota$ is the binary inclination.

The phases of the harmonics follow a simple relation $\frac{1}{m}\Psi^\mathrm{SPA}_{\ell m} (m f) = \frac{1}{2}\Psi^\mathrm{SPA}_{22} (2f)$ \cite{Mis16}.
However, these SPA-based formulae can have significant deviations for high-mass binaries, for which the non-linear regime is within the band of the detectors. We thus use the waveform approximant \texttt{IMRPhenomXHM} instead of Eq.~\ref{eq:AmpFormulae} to construct our templates \cite{Gar20}. Note that we do not consider the modes beyond $(4, 4)$ as the SNR in these modes progressively decreases, and also because they are not currently modeled in the \texttt{IMRPhenomXHM} approximant \cite{Gar20}.

We separately store normalized templates for individual harmonics, which are given by
\be
\mathbb{h}_{\ell m}(f) \equiv A^\mathrm{bank}_{\ell m}(f) e^{i \Psi^\mathrm{model}_{\ell m}(f)}
\label{eq:NormTemplates}
\ee
where $A_{\ell m}^\mathrm{bank} (f)$ corresponds to the average of normalized waveforms (i.e., $\langle A_{\ell m}^\mathrm{bank} (f)| A_{\ell m}^\mathrm{bank}(f)\rangle=1$) over the parameter range corresponding to each bank. We model the phases of the templates as,
\be
\Psi^\mathrm{model}_{22} (f) \equiv \langle\Psi_{22}\rangle_\mathrm{subbank} (f) + \sum_{i=0}^{10} c^{(22)}_i \Psi^\mathrm{SVD}_i (f),
\label{eq:phases_22}
\ee
where $\Psi_i$ corresponds to the orthonormal basis functions from the singular value decomposition (SVD), and $\langle \Psi \rangle_\mathrm{subbank}$ denotes the average over physical waveforms in the particular subbank. We use the random forest regressor (RF) to model: $\{c^{(22)}_2, \ldots , c^{(22)}_{10} \} = \mathrm{RF} (c_0^{(22)}, c_1^{(22)})$. Thus, all our template waveforms correspond to grid points in a two dimensional space of $(c_0^{(22)}, c_1^{(22)})$. We use the implementation of the RF regressor from the \texttt{scikit-learn} package~\cite{scikit_learn}. We use the following model for the (3,3) phases:

\be\begin{split}
\Psi^\mathrm{model}_{33}(f) \equiv  \frac{3}{2}& \Psi_{22}\left(\frac{2f}{3}\right) + \langle\Delta \Psi_{33}\rangle_\mathrm{bank} (f)\\
 +& \sum_{i=0}^{2} \mathrm{RF}_i (c_0^{(22)}, c_1^{(22)}) \Psi^\mathrm{SVD}_i(f) 
\label{eq:33_phase_model}.
\end{split}\ee
and a similar one for (4,4).
The first term on the RHS corresponds to the SPA relation, while the others are obtained by decomposing the residuals obtained from \texttt{IMRPhenomXHM} into a mean term and modeling the remaining variation using a few leading SVD vectors as
\be\begin{split}
\Psi^\mathrm{XHM}_{33}&(f)-\frac{3}{2} \Psi^\mathrm{XHM}_{22}\left(\frac{2f}{3}\right) \\
&\simeq \langle\Delta \Psi_{33}\rangle_\mathrm{bank} (f)  + \sum_{i=0}^{2} c^{(33)}_i \Psi^\mathrm{SVD}_i(f) 
\end{split}\label{eq:HM_phase_residuals}\ee
We could have added all the $c^{(33)}$ dimensions to our template grid, however, this can lead to a large increase in the number of templates. We therefore use RF to model the leading $c^{(33)}_i$ as a function of the leading $(2,2)$ coefficients in Eq.~\eqref{eq:33_phase_model}.

As we use normalized amplitudes for each harmonic template, we need extra factors to account for the relative amplitudes of the harmonics. For example, we expect high-mass and asymmetric mass-ratio binaries to emit relatively higher amplitudes of HM and vice versa. We therefore include samples of physically possible fractional SNR ratios in HM given by
\be
R_{\ell m} \equiv \frac{\langle h^\mathrm{XHM}_{\ell m}(f)|h^\mathrm{XHM}_{\ell m}(f)\rangle^{1/2}}{\langle h^\mathrm{XHM}_{22}(f)|h^\mathrm{XHM}_{22}(f)\rangle^{1/2}}
\label{eq:Rlm}\ee
which are calculated at $\iota=\pi/2$.
We also include a weight for each sample corresponding to observable volume at a reference binary inclination ($w\propto \langle h^\mathrm{XHM}|h^\mathrm{XHM}\rangle^{3/2}$, where $h^\mathrm{XHM}$ has all the modes included in it). For example, we give a larger weight to systems with large masses and symmetric mass ratios, as they are observable to a larger distance, and vice versa.
We also normalize the weights such that $\sum_i w_i=1$.
$R_{\ell m}$ are scalars which vary with masses and spins of waveforms.

In an individual detector, a physical waveform (with an arbitrary overall normalization) can thus be written as a linear combination of our templates as
\be\begin{split}
h(f)=&\, e^{i2\phi_0} \h_{22}+ R_{33} \sin \iota\, e^{i3\phi_0}  \h_{33}+ R_{44} \sin^2 \iota\, e^{i4\phi_0} \h_{44}
\end{split}\label{eq:full_wf_bank}
\ee
where again $\iota$ is the inclination of the binary, and $\phi_0$ is the initial reference phase. For each bank, we store a separate set of $R_{\ell m}$ samples (for reference, see the blue points in the left panel of Fig.~\ref{fig:Mode_ratios}), which we will later use in section~\ref{sec:NewDetStats} as our prior expectation for physically possible SNRs in different modes for a given inclination. 

\subsection{Matched filtering procedure}

In principle, we can filter the data separately with each of our harmonic templates from Eq.~\eqref{eq:NormTemplates} and collect the complex SNR timeseries given by
\be \begin{split}
\rho_{\ell\ell} (t) &\equiv \langle \h_{\ell\ell}(f)\ |\ d(f)\ e^{i2\pi f t}\rangle
\label{eq:SNR_timeseries_unorth}
\end{split}\ee

A slight complication in such an analysis is that the harmonics are not exactly orthogonal.
The off-diagonal terms of their covariance matrix $C_{\ell \ell'} \equiv \langle \mathbb{h}_{\ell\ell}\, (f)|\mathbb{h}_{\ell'\ell'}\, (f)\rangle$ are typically $\lesssim 0.1$, but at times can be larger (e.g., in the case of short duration waveforms for particular cases of PSD profiles).
Their inner product to calculate log likelihood is given by $(\sum_{\ell\ell'}\rho^*_{\ell\ell} [C^{-1}]_{\ell \ell'} \rho_{\ell'\ell'})^{1/2}$.

We however choose an alternative path to simplify our calculation during the matched-filtering stage. We orthogonalize the harmonics using the Gram--Schmidt method: we keep (2,2) fixed and orthogonalize (3,3) with respect to (2,2), and then orthogonalize (4,4) with respect to (2,2) and the orthogonalized (3,3).
We then normalize the harmonics such that  $ \langle \mathbb{h}^\perp_{\ell\ell}\, (f)|\mathbb{h}^\perp_{\ell'\ell'}\, (f)\rangle = \delta^K_{\ell \ell'}$, where $\delta^K$ is the Kronecker delta function and record the SNR timeseries as
\be \begin{split}
\rho^\perp_{\ell\ell} (t) &= \langle \h^\perp_{\ell\ell}(f)\ |\ d(f)\ e^{i2\pi f t}\rangle
\label{eq:SNR_timeseries}
\end{split}\ee
Note that $\rho^\perp_{22} = \rho_{22}$ by construction. The waveforms in the two cases are related as $\h^\perp_{\ell \ell} = L_{\ell \ell'} \h_{\ell' \ell'}$, where $L$ is the lower triangular matrix obtained using Cholesky decomposition of the inverse of the mode covariance matrix  $C^{-1} = L L^\dagger$. The SNR amplitudes are also similarly related as $\rho^\perp_{\ell \ell} = L_{\ell \ell'} \rho_{\ell' \ell'}$ and we use these two cases in different scenarios in this paper.

\section{Trigger-ranking statistics with HM}
\label{sec:NewDetStats}

In this section, we derive ways to optimally combine the SNR timeseries from different modes.
In a $(2, 2)$-only search, we require $|\rho_{22}|^2$ to be above a certain threshold in order to collect triggers in individual detectors.
Naively, one could use the following detection statistic in the case with HM
\be
\rho^2_\mathrm{HM} = |\rho^\perp_{22}|^2+|\rho^\perp_{33}|^2+|\rho^\perp_{44}|^2
\label{eq:Match}
\ee

Under the Gaussian noise hypothesis, this will follow a $\chi^2$-distribution with 6 degrees of freedom (corresponding to the real and imaginary parts of each of the modes). We show in orange points in Fig.~\ref{fig:Mode_ratios} the triggers (mostly contributed by Gaussian noise) which are allowed upon making a cut with $ |\rho^\perp_{22}|^2+|\rho^\perp_{33}|^2+|\rho^\perp_{44}|^2>25$ (a minor detail to note is that we show triggers maximized over intervals of 0.01\,s, to avoid showing correlated triggers at very small time separations). We also show as blue points the parameter space covered by physical injections from \texttt{IMRPhenomXHM}.
We can immediately see that only a small subset of the mode ratio parameter space is populated by physical injections. This is because the $(3,3)$ and $(4,4)$ SNR amplitudes are strongly correlated as they both peak for edge-on asymmetric mass-ratio systems. The statistic $\rho^2_\mathrm{HM}$ in Eq.~\eqref{eq:Match} allows for triggers with arbitrarily large mode ratios, and thus using a simple cut on its value will generate many unphysical triggers. Notice that $\rho^2_\mathrm{HM}$ was motivated entirely for the case of the noise hypothesis, but we will therefore construct a more optimal detection statistic that uses the knowledge of both the signal and noise hypothesis.

We start with Neyman--Pearson lemma \cite{neymanpearson}, which states that the optimal detection statistic is the ratio of evidence under the signal ($\mathcal{S}$) and the noise ($\mathcal{N}$) hypothesis:
\be\begin{split}
\exp\left(\frac{\rho^2_\mathrm{detection}}{2}\right) \equiv \frac{P(d|\mathcal{S})}{P(d|\mathcal{N})} = & \frac{P(d|\mathcal{S})}{P(d|\mathrm{GN})} \frac{P(d|\mathrm{GN})}{P(d|\mathcal{N})}
\end{split}\label{eq:NeymanPearson}\ee
We deliberately break the statistic into two parts and denote the hypothesis when the noise is Gaussian by GN.
We denote the data in detector $k$ by $d_k$ and our model waveform by $h_k$. $h_k$ is a function of the intrinsic binary parameters $\theta_\mathrm{int} \in \{m_1, m_2, s_{1z}, s_{2z}\}$ (we have assumed only the aligned spin components in our search) and extrinsic parameters $\theta_\mathrm{ext} \in \{\iota, \phi_0, D, \psi, \hat{\bm{n}}\}$, which correspond to inclination, initial orbital phase, luminosity distance polarization and sky position (right ascension and declination) respectively. The first term corresponding to Gaussian noise can be written as

\begin{widetext}

\be\begin{split}
&\frac{P(d|\mathcal{S})}{P(d|\mathrm{GN})}  =  \int d\Pi(\theta_\mathrm{int},\theta_\mathrm{ext}) \exp \left[ \sum_{k\in \mathrm{detectors}} \frac{1}{2}\langle d_k | d_k \rangle - \frac{1}{2}\langle d_k-h_k | d_k-h_k\rangle \right]\\
= & \int d\Pi(\theta_\mathrm{int},\theta_\mathrm{ext}) \exp \left[ \sum_{k\in \mathrm{detectors}} \mathrm{Re}(\langle h_k | d_k \rangle) - \frac{1}{2}\langle h_k | h_k \rangle \right]\\
= & \sum_{\alpha \in \mathrm{templates}} P(\alpha|\mathcal{S}) \bigg \{ \int d\Pi(R_{33},R_{44},\theta_\mathrm{ext}|\alpha) \exp \left[ \sum_{k\in \mathrm{detectors}} \mathrm{Re}(\langle h_k | d_k \rangle) - \frac{1}{2}\langle h_k | h_k \rangle \right] \bigg \}\\
\equiv & \sum_{\alpha \in \mathrm{templates}} P(\alpha|\mathcal{S})\, e^{\rho_\mathrm{coherent}^2 (\alpha)/2},
\label{eq:CoherentScore}
\end{split}
\ee
where we sum over the contribution from different templates. $P(\alpha|\mathcal{S})$ is the astrophysical prior corresponding to the template $\alpha$, $\Pi$ is the prior density over mode amplitude ratios and extrinsic parameters. $R_{\ell m}$ are the physical mode SNR ratios defined in Eq.~\eqref{eq:Rlm}. We denote the logarithm of the expression in the curly brackets as our Gaussian coherent score $(\rho_\mathrm{coherent}^2)$ corresponding to a given template. This is calculated using the marginalization algorithm implemented in the publicly available \texttt{cogwheel}\footnote{\url{https://github.com/jroulet/cogwheel/tree/v1.2.1}} package \cite{Rou22_cogwheel,Rou23_CoherentScore} (see the subsection~\ref{sec:CoherentScore} below for details).
We calculate the second quantity in the RHS of Eq.~\eqref{eq:NeymanPearson} related to the non-Gaussian correction as
\be\begin{split}
\frac{P(d|\mathrm{GN},\alpha)}{P(d|\mathcal{N},\alpha)}
\simeq \exp\bigg[  \sum_{k\in \mathrm{detectors}} -\frac{|\rho_k|^2}{2} - \log (P(|\rho_k|^2 \mid \alpha, \mathcal{N}))\bigg]
\equiv e^{ -\sum_k \Delta \rho^2_k(\alpha)/2},
\end{split}\label{eq:NG_correction}\ee
where the probability $P(|\rho_k|^2 \mid \alpha, \mathcal{N})$ is computed using the empirical distribution of triggers obtained from timeslides for a given template $\alpha$, and we use $|\rho|^2 \equiv \rho_{\ell \ell} C^{-1}_{\ell \ell'} \rho_{\ell'\ell'}$ is the effective SNR after taking into account the relative covariance between modes. Our final trigger-ranking statistic is given by
\be\begin{split}
 \exp\left(\frac{\rho^2_\mathrm{ranking}}{2}\right) = & \sum_{\alpha \in \mathrm{templates}} P(\alpha|\mathcal{S})\, e^{\frac{1}{2}\rho_\mathrm{coherent}^2 (\alpha) - \frac{1}{2}\sum_{k} \Delta \rho_k^2 (\alpha)}\\
 \simeq &\, P(\alpha_\mathrm{max}|\mathcal{S})\, e^{\frac{1}{2}\rho_\mathrm{coherent}^2 (\alpha_\mathrm{max})- \frac{1}{2}\sum_{k} \Delta \rho_{k}^2 (\alpha_\mathrm{max})}
 \end{split}\label{eq:full_det_statistic}\ee
 Ideally, one should marginalize over both intrinsic and extrinsic parameters, however this becomes too expensive to perform. Therefore, in the current version, we use an approximation which maximizes over intrinsic parameters. This corresponds to picking the (2,2) template ($\alpha_\mathrm{max}$) which gives the maximum value inside the sum in Eq.~\ref{eq:full_det_statistic}. In the future, we hope to relax this assumption and calculate the integral over also the intrinsic parameters. We discuss the calculation of the Gaussian coherent score and the non-Gaussian correction separately in the following two subsections.

\subsection{Coherent score in Gaussian noise}
\label{sec:CoherentScore}
Using the formula in Eq.~\eqref{eq:full_wf_bank} for a particular template from the template bank, the predicted waveform $h_k$ of the signal in detector $k$ becomes
\be \begin{split}\label{eq:predictedwf}
  h_k (f) \simeq\, \frac{D_0}{D}\left[ F_{+, k} \frac{1 + \cos^2 \iota}{2} \, - \, i\cos \iota\, F_{\times, k}\right] \bigg[ & e^{2 i \phi_0} \,  \mathbb{h}_{22} (f) + \, e^{3 i \phi_0} \, R_{33}(M,q)\ \sin \iota\ \mathbb{h}_{33} (f)\\ &+ \, e^{4 i \phi_0} \ R_{44}(M,q)\ \sin^2 \iota\ \mathbb{h}_{44}(f)\bigg]\, ,
\end{split}\ee
where $\mathbb{h}$ corresponds to complexified unit templates ($\rm SNR=1$) for different modes and $D_0$ is the distance where $\rm SNR=1$ for the $(2,2)$ mode. $F_{p, k}$ are the antenna response of the detector to polarization $p \in \{+,\times\}$ and is dependent on the polarization angle $(\psi)$ and sky location $(\hat{\bm{n}})$.  $R_{\ell m}$ samples correspond to variation of the the ratio of the SNR in the $(\ell, m)$ mode to that in $(2, 2)$ at $\iota=\pi/2$ over intrinsic parameters in the subbank, see Eq.~\eqref{eq:Rlm}. 

In our analysis, we only consider the $\ell=m$ modes, for which the factorization of the polarization terms in Eq.~\eqref{eq:predictedwf} is possible \cite{Mil21}.
Let us now calculate the inner products $\langle h|d\rangle$ and $\langle h|h\rangle$ used in Eq.~\eqref{eq:CoherentScore}.
For $\langle h|h\rangle$,
\be \begin{split}
  \langle h (f) | h (f) \rangle =& \sum_{\ell, \ell'} \langle
  h_{\ell \ell}(f) | h_{\ell'\ell'}(f) \rangle = \sum_{\ell, \ell'}  \frac{D^2_0}{D^2}\bigg| F_{+} \frac{1 + \cos^2 \iota}{2} \, - \, i\cos \iota\, F_{\times}\bigg|^2 e^{i\phi_0 (\ell'-\ell)}\,  \mathbb{C}_{\ell,\ell'}
\end{split}\label{eq:hh}\ee
where the covariance matrix between the modes is given by
\be \begin{split}
\mathbb{C} = \begin{bmatrix}
1 & R_{33} \sin \iota\, \langle\h_{22}(f)|\h_{33}(f)\rangle & R_{44} \sin^2 \iota\, \langle\h_{22}(f)|\h_{44}(f)\rangle\\
- & R_{33}^2 \sin^2 \iota & R_{33}R_{44} \sin^3\iota\, \langle\h_{33}(f)|\h_{44}(f)\rangle\\
- & - & R_{44}^2 \sin^4 \iota
\end{bmatrix}
\end{split}\label{eq:covariance_modes}\ee
(the lower-triangular elements are not shown but can be obtained by Hermitian conjugation). For $\langle h|d\rangle$,
\be \begin{split}
\langle h (f) | d_k (f)& e^{i2\pi f t}  \rangle = \sum_{\ell} \langle
  h_{\ell \ell} (f) | d_k (f) e^{i2\pi f t} \rangle\\
  =& \frac{D_0}{D}\left[ F_{+, k} \frac{1 + \cos^2 \iota}{2} \, + \, i\cos \iota\, F_{\times, k}\right] \bigg[e^{2i\phi_0} \rho_{22,k}(t) + e^{3i\phi_0} R_{33} \sin \iota\, \rho_{33,k}(t) + e^{4i\phi_0} R_{44} \sin^2 \iota\, \rho_{44,k}(t)\bigg],
\end{split} \label{eq:dh} \ee
where we have denoted the complex inner product of the data with the unit template in detector $k$ as $\rho_{\ell\ell,k}(t)$. The expected time of arrival of signal in each detector is given by $t_k \equiv t_\oplus + \bm{r}_k\cdot \hat{\bm{n}}/c$, which is the arrival time at the geocenter plus a correction depending on the location of detector $\bm{r}_d$. We compute Eq.~\eqref{eq:CoherentScore} for given a template $\alpha$ as
\be
e^{\rho_\mathrm{coherent}^2/2}  \simeq \sum_{i}w^{(i)}
 \int d\Pi(\iota, \phi_0, D, \psi, \hat{\bm{n}}) \exp \left[ \sum_{k\in \mathrm{detectors}} \ln \mathcal{L}_k (t_k, R^{(i)}_{33}, R^{(i)}_{44}, \iota, \phi_0, D, \psi, \hat{\bm{n}}) \right]
\label{eq:MultiDetCS}
\ee
\end{widetext}
where the log-likelihood is $\ln \mathcal{L}_k \equiv \mathrm{Re}(\langle h_k | d_k \rangle) - \frac{1}{2}\langle h_k | h_k \rangle$, and we calculate the marginalization over $D, \phi_0$, $\hat{\bm{n}}, \psi, t_\oplus$ and $R_{\ell m}$. Let us first focus on calculating the integral over mode amplitude ratios. As we show in Eq.~\ref{eq:MultiDetCS}, we use Monte Carlo integration using the pre-generated $R_{\ell m}$ samples in Eq.~\eqref{eq:Rlm}.
$w$ is again the weight of each sample corresponding to its observable volume.
Currently, we assume each template in a given subbank has the same set of mode ratio samples. One could however use template-dependent mode ratio samples by generating them using normalizing flows, we leave this approach to an upcoming work.

Algorithms to calculate each of the remaining integrals are discussed in detail in \citet{Rou23_CoherentScore}. We integrate over $D$ by interpolating a precomputed table, over $\phi_0$ by trapezoid quadrature, and over the remaining extrinsic parameters $\iota, \hat{\bm{n}}, \psi$ using adaptive importance sampling. Overall, given a multi-detector trigger, our pipeline currently need $\sim 200$ ms for calculating the full marginalization integral. We provide an example of the marginalized integral calculation in a jupyter notebook \nblink.

It is worth noting that in a search pipeline, before we rank the triggers to calculate their false alarm rate, we need to first collect triggers by filtering the templates with strain from individual detectors. However, in that stage, we do not have information on the triggers in the other detectors, and cannot directly use Eq.~\eqref{eq:MultiDetCS}. We therefore generalize the results in this section in Appendix~\ref{sec:single_det_statistic} and use Eq.~\eqref{eq:SingleDetScore} for collecting triggers from individual detectors.


Note that in the IAS $(2, 2)$-only search, our coherent score was also computed as the integral over the extrinsic parameters (see Appendix D of \cite{Ols22_ias_o3a}). Here we presented a generalized version of the formalism for the case of aligned-spin HM.



\subsection{Non-Gaussian noise correction to the detection statistic}
\label{sec:nonGaussian_statistic}

\begin{figure}
\centering
\includegraphics[scale=0.7,keepaspectratio=true]{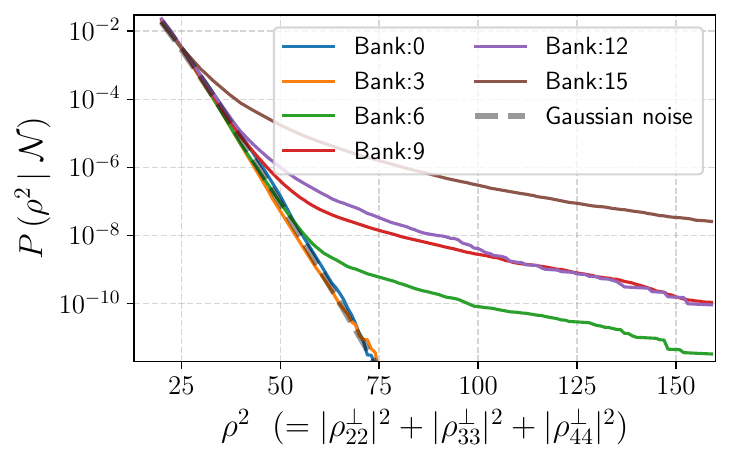}
\includegraphics[scale=0.7,keepaspectratio=true]{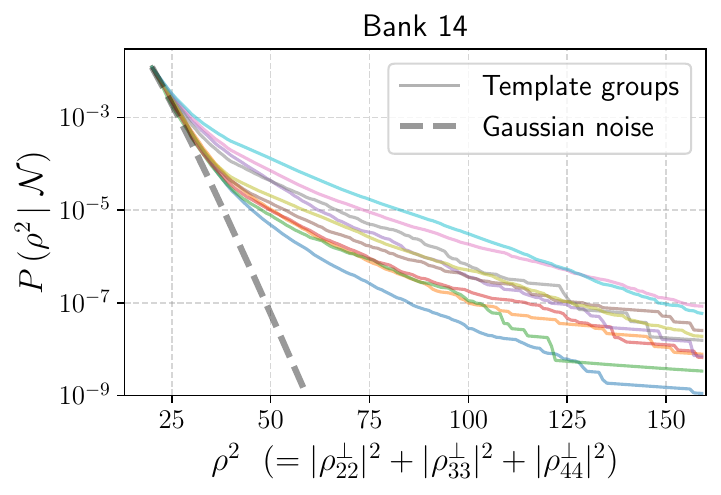}
\caption{In our trigger-ranking statistic, we incorporate the probability of a trigger with a given SNR occurring under the noise hypothesis (see Eq.~\eqref{eq:NG_correction}). To infer this, we construct empirical probability functions using the histogram of background triggers obtained from timeslides. \textbf{Top:} For our banks, mass of the binary roughly increases with bank \# and vice versa for the waveform duration. Our higher banks are indeed relatively much more dominated by short-duration non-Gaussian transients (glitches). For the very high mass banks ($M_\mathrm{tot}>200 M_\odot$) we see non-Gaussianity dominating even at low SNRs ($\rho^2 \sim 40$). \textbf{Bottom:} Instead of using a common $P(\rho^2,\mathcal{N})$ curve for all templates in a subbank, we construct separate probability functions for different groups of templates (shown as colored lines). For one of the very high-mass banks (BBH-14,0), we clearly see that high-SNR triggers have a very different chance of occurring due to noise for different templates and we exploit this fact in our trigger-ranking statistic.}
\label{fig:RankFunction}
\end{figure}

Previously, in Eqs.~\eqref{eq:NG_correction} and~\eqref{eq:full_det_statistic}, we had introduced a term $e^{\frac{1}{2}\Delta \rho^2 (\alpha)}$ to account for the correction due to non-Gaussian nature of noise in a particular detector. In our case of combining scores from the three (orthogonalized) modes:
\be\begin{split}
e^{-\frac{1}{2}\Delta\rho^2 (\alpha)} \simeq & \exp\bigg[-\frac{|\rho^\perp_{22}|^2+|\rho^\perp_{33}|^2+|\rho^\perp_{44}|^2}{2}\\
&- \log [P_{\alpha}(|\rho^\perp_{22}|^2,|\rho^\perp_{33}|^2, |\rho^\perp_{44}|^2 | \mathcal{N})]\bigg]
\end{split}\label{eq:p_alpha}\ee
where $P_{\alpha}(|\rho^\perp_{22}|^2,|\rho^\perp_{33}|^2, |\rho^\perp_{44}|^2 | \mathcal{N})$ is the probability of getting a set of $\{\rho^\perp_{22}, \rho^\perp_{33}, \rho^\perp_{44}\}$ from a noise trigger for a template $\alpha$. Note that under Gaussian noise, $e^{-\Delta\rho^2 (\alpha)}$ should be $\sim 1$.
However, in a realistic scenario, its value can be lower by multiple orders of magnitude. We currently do not have a good theoretical model for the distribution of the non-Gaussian noise, hence we need to empirically estimate $P_{\alpha}$ using the histogram of the number of background triggers (obtained from timeslides) caught by a template $\alpha$:
\be\begin{split}
P_{\alpha}(|\rho^\perp_{22}|^2,|\rho^\perp_{33}|^2,& |\rho^\perp_{44}|^2 | \mathcal{N})\\ 
&\propto \left [ \frac{d N_\alpha}{d (|\rho^\perp_{22}|^2) d (|\rho^\perp_{33}|^2) d (|\rho^\perp_{44}|^2)} \right ]^{-1}
\end{split}\ee
It is worth noting for reference that, in previous (2,2)-only IAS pipeline papers, the quantity $-2\log P$ was termed as the ``rank function'' (cf. section~J of Ref.~\cite{ias_pipeline_o1_catalog_new_search_prd2019}).

Estimating the three-dimensional $P_{\alpha}$ distribution from a finite number of timeslides can lead into sparsity issues especially at high SNR values. However, we cannot arbitrarily increase the number of timeslides as the cost of search also increases. Therefore, to ameliorate the sparsity issue, instead of making a 3D histogram, we will construct a 1D histogram of the quantity $\rho^2_\mathrm{sum}\equiv |\rho^\perp_{22}|^2 + |\rho^\perp_{33}|^2 + |\rho^\perp_{44}|^2$ using background triggers. We then use the approximation 
\be
P^\mathrm{3D}_{\alpha}(|\rho^\perp_{22}|^2, |\rho^\perp_{33}|^2, |\rho^\perp_{44}|^2| \mathcal{N}) \simeq \frac{16}{\rho^4_\mathrm{sum}} P^\mathrm{1D}_{\alpha}(\rho^2_\mathrm{sum}| \mathcal{N})
\label{eq:1D_background}\ee
The constant of proportionality in the above equation is based on the following behavior in the case of Gaussian noise. $\rho^2_\mathrm{sum}$ follows a $\chi^2$-distribution with 6 degrees of freedom (corresponding to the real and imaginary parts of each modes) and $P_{\chi^2}(\rho^2;6) = \frac{\rho^4}{16} e^{-\rho^2}$. On the other hand, $P_{\alpha}(|\rho^\perp_{22}|^2, |\rho^\perp_{33}|^2, |\rho^\perp_{44}|^2| \mathcal{N})$ is the product of three independent $\chi^2$-distributions each with 2 degrees of freedom and $P_{\chi^2}(\rho^2;2) = e^{-\rho^2}$.

Another possibility to simplify the calculation of $P_{\alpha}$ in Eq.~\eqref{eq:p_alpha} would have been to assume independent SNR distributions of the modes: $P_{\alpha}(|\rho^\perp_{22}|^2, |\rho^\perp_{33}|^2, |\rho^\perp_{44}|^2| \mathcal{N}) \propto P_{\alpha}(|\rho^\perp_{22}|^2| \mathcal{N}) P_{\alpha}(|\rho^\perp_{33}|^2| \mathcal{N}) P_{\alpha}(|\rho^\perp_{44}|^2| \mathcal{N})$. However, this approximation may be inaccurate as glitches can trigger multiple modes at once and correlate their SNRs. 

We show the 1D probability function from Eq.~\eqref{eq:1D_background} for different banks in Fig.~\ref{fig:RankFunction}. The bank number roughly increases with the total binary mass. The high mass banks have short duration template waveforms and are thus more susceptible to short-duration transients produced by instrumental glitches \cite{LIGO_O1, CoherentScore, psd_drift, BlipGlitches}. The overall normalization of all the curves has been arbitrarily adjusted so that they match at $\rho^2=20$.

In the previous IAS studies, we used the same probability function $P_{\alpha}(|\rho^\perp_{22}|^2)$ for triggers from all templates in a subbank. However, even within a particular bank, different templates can have significantly different susceptibility to glitches (especially in the high mass case).
To showcase this effect, we empirically divide the templates into different groups based on the fraction of triggers above a certain reference SNR: $N(\rho^2>\rho^2_\mathrm{ref})/N(\rho^2\leq\rho^2_\mathrm{ref})$. We pick a high reference SNR (typically $\rho^2\gtrsim 60$) such that it is beyond the Gaussian dominated regime for the particular subbank. We show the 1D probability function for different template groups in a particular high mass bank ($M_\mathrm{tot}\gtrsim 200\, M_\odot$) in the lower panel of Fig.~\ref{fig:RankFunction}. We can infer that using the same probability function for all templates is less optimal as different templates tend to catch significantly different number of glitches (and hence there should be a difference in the significance of the triggers obtained from different templates). A minor detail to note is that we make template groups such that each group has $\gtrsim 500$ background triggers to avoid the problem of sparsity in our construction of the probability function.


\section{Data preprocessing}
\label{sec:DataPreproc}

Before we collect triggers by matched-filtering templates against the strain data, we identify and excise bad data segments localized in time (i.e., make ``holes'') and then inpaint these segments (see sections~C and D of \cite{ias_pipeline_o1_catalog_new_search_prd2019} and also~\cite{psd_drift, vetopaper}). For a particular time-segment to be excised, we use the criterion that the power registered in certain pre-defined time-frequency regions (e.g., [20--50]$\rm\,Hz \times 1\,s$) exceeds particular thresholds. To determine the threshold corresponding to a particular time-frequency region, we choose a few reference templates in the bank, rescale them to a high SNR value (in our case, $\rm SNR=20$) and then calculate the power falling inside that time-frequency region. We store the highest registered power value as the threshold power for that particular time-frequency region and remove transients in real data which cross this threshold. In this way, we aim to guarantee that astrophysical signals with SNR below 20 are preserved.

Note that the thresholds are calculated corresponding to the reference bank templates and are thus different for different banks. In the earlier IAS pipeline papers, we only used (2,2) templates to calculate this threshold. However, if there is a high SNR physical signal which includes HM, there is a possibility we could excise that signal if we only use (2,2) templates. Hence, we now use the full templates including HM (from Eq.~\eqref{eq:full_wf_bank}) to calculate the threshold more accurately. 


\subsection{Erasing time-varying noisy bands with\\ band eraser}
\label{sec:Band_eraser}

\begin{figure*}
\centering
\includegraphics[width= 1.\textwidth,keepaspectratio=true]{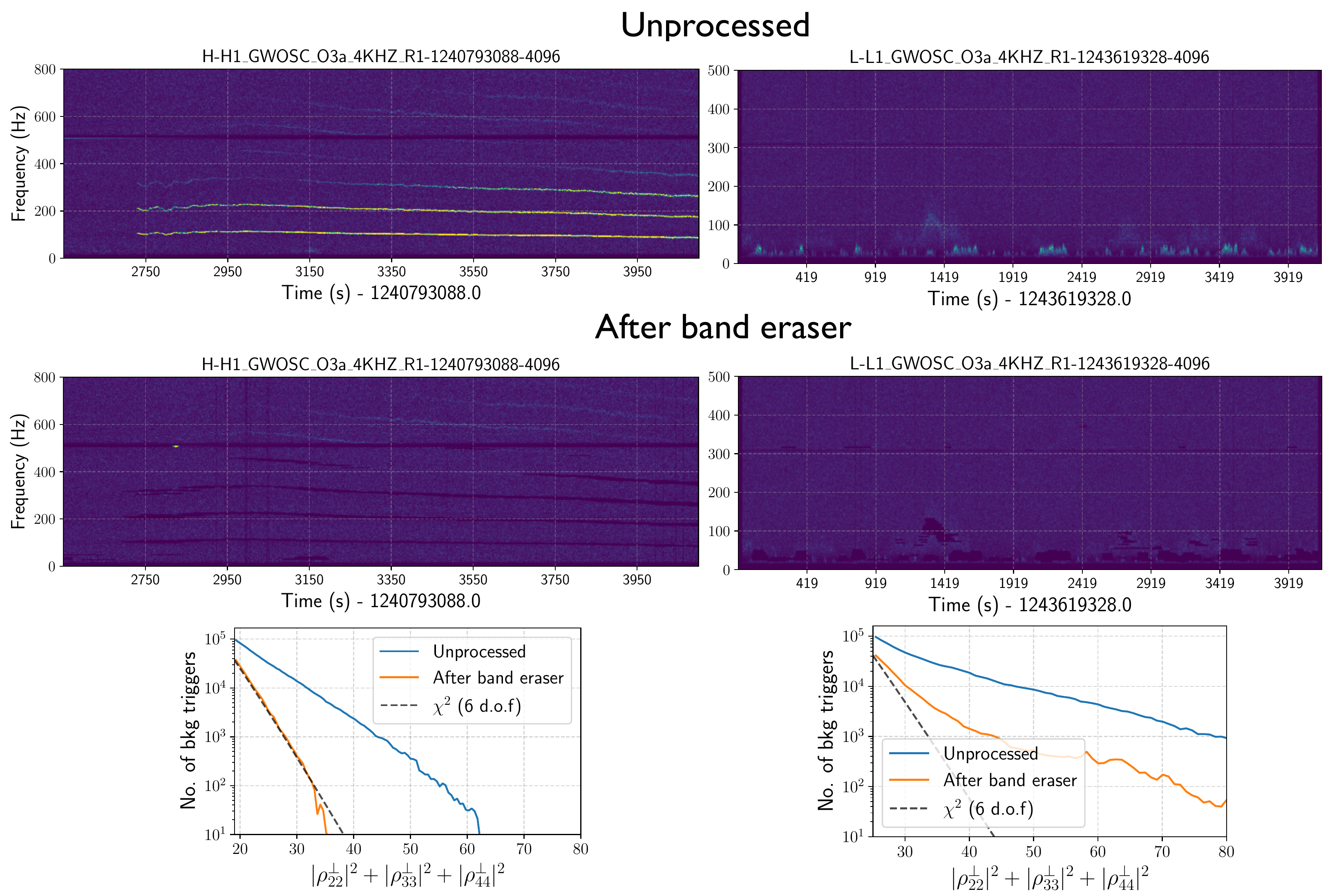}
\caption{\textbf{Top:} We show two examples of the worst Hanford and Livingston unprocessed O3a files in terms of the number of glitches caught by one of our high mass banks in our pipeline testing phase. \textbf{Middle:} We therefore develop a tool called ``band-eraser" which we use to remove time-varying bands with excess power in spectrograms (see section~\ref{sec:Band_eraser}). Application of the tool to the files gives the cleaned spectrogram shown. \textbf{Bottom:} We show the relative improvement in the number of background triggers from individual files. Note also that the band eraser also prevents unnecessary excision of entire time-chunks if the disturbances are localized to a small range of frequencies. The improvement in the overall sensitive volume-time due to band eraser will be explored in a future paper. A test to measure the safety of the band eraser using a low-mass physical injection in these files is shown in Fig.~\ref{fig:BE_injection}. \nblink}
\label{fig:BE_specgram}
\end{figure*}

One of the downsides of excising entire time chunks of data is that information is removed from all frequency bands.
Sometimes, however, the noise transients are only localized to specific frequency bands (particularly the low-frequency end $\lesssim$ 100 Hz) and do not affect astrophysical signals which dominantly have power in a different frequency band. We show two examples of such cases in the upper panel of Fig.~\ref{fig:BE_specgram}. These examples produced the largest number of background triggers in our high-mass banks (where the templates have a large power at low frequencies) during testing phase of our pipeline. In numerous other cases also we noticed that the noisy regions were localized to low frequencies, some of which can be attributed to known disturbances like scattered light \cite{Son21_Scattered_Light}.

We therefore develop a new tool, ``band eraser", to smoothly remove bad time-frequency regions (or bands) in the spectrogram. We first divide the short-time Fourier transform (obtained using \texttt{scipy.signal.stft}) of whitened data into ``bands", i.e., segments with dimensions $\rm 64\,s \times 2\,Hz$. We further split each band into chunks of size $\rm 2\,s \times 0.5\,Hz$ and calculate the number of chunks in each band with power greater than a reference threshold. We remove the entire band from our analysis if the number of chunks is above a certain safety threshold (we calculate the safety threshold using Poisson statistics, making sure that, on average, in Gaussian noise, one band is removed per $\sim 4 \times 10^4$\,s). The removal of noise is done by multiplying the signal in the flagged time-frequency band by zeros and then using \texttt{scipy.signal.istft} to obtain the cleaned time-domain strain.
In the middle panel of Fig.~\ref{fig:BE_specgram}, we show spectrograms of the same files as the top panels but with the band eraser applied (the dark stripes correspond to flagged time-frequency bands which were zeroed out). In the bottom panel, we show the histogram of the triggers registered from the unprocessed and cleaned files and we see a significant reduction in the number of non-Gaussian triggers (the expected histogram from Gaussian noise is shown in the dashed black line corresponding to 6 degrees of freedom, corresponding to real and imaginary part of the three harmonics).

After the band eraser is applied, we perform the operations mentioned earlier: make holes to remove the remaining disturbances in the data and then inpaint the holes.
 
We also checked that physical injections corresponding to binary black holes are not removed due to the band eraser. We inject a particular low-mass waveform in the files shown in Fig.~\ref{fig:BE_specgram} and then show the result from using band eraser close to the injection in Fig.~\ref{fig:BE_injection}.
We plan to do injections in the entire O3 run to measure the sensitivity of our pipeline in a future study. The sensitive volume time $VT$ could slightly increase upon using the band eraser due to two effects: reduction in the noise background, and salvaging of the data in regions cleaned by the band eraser (as there is no more need to excise entire time chunks). Erasing noisy bands could potentially also improve the stability of the parameter estimation runs \cite{Cor15_BayesWave}. We also provide an example jupyter notebook
 for the band eraser code \nblink.

\section{Discussion}
\label{sec:Discussion}

\subsection{Applicability of the mode-by-mode filtering method to other search pipelines}
We implemented the mode-by-mode filtering method in the IAS pipeline, but we believe the formalism that we develop in this paper is general and can be implemented to search for HM in other template-based pipelines in the literature, e.g., \cite{gstlal, PYCBCPipeline, mbta_o3a_pastro_andres2022, nitz_4ogc_o3_ab_catalog_2021}. For using the mode-by-mode filtering method, all one needs is normalized templates for different harmonics ($\h_{22}, \h_{33}, \h_{44}$) and samples for physical mode SNR ratios ($R^{(i)}_{33}, R^{(i)}_{44}$) associated with the templates. In section~\ref{sec:TemplateBanks}, we gave an overview of adding HM templates to a $(2,2)$-only bank constructed using geometric placement. We discuss in our companion paper~\cite{Wad23_TemplateBanks} methods to add HM templates to banks constructed using stochastic placement. It is also worth mentioning a different example where mode-by-mode filtering was applied in a $(2,2)$-only search to reduce the matched-filtering cost: \cite{Mes17_GstLAL_LowLatency} (where filtering with data was done separately with different SVD components of the $(2,2)$ waveform).

\subsection{Dependence on astrophysical prior}
Similar to the $(2,2)$-only search, we assume an astrophysical prior to construct template banks for our search. The search including HM can lose sensitivity if the assumed astrophysical distribution is different from the true one.
One option to overcome this issue is to reweight the results from the search (IFAR and $p_\mathrm{astro}$ values of triggers) in a hierarchical population analysis, see e.g., \cite{ias_popO2_Roulet_2020}.
Note that a better knowledge of the distribution of masses and mass ratios upon detecting more events with future observatories (\cite{Ng22_ET_HM, Fai23, einsteinTelescope_Maggiore:2019uih, Eva23, decigo_spaceDetector_kawamura2020, LISA, KAGRA:2020tym}) will also reduce the sensitivity loss in the search resulting from a wrong assumption of the astrophysical prior. To estimate how much exactly is the gain or loss in our overall volume-time ($VT$) sensitivity due to addition of HM in templates, we aim to perform an injection study in an upcoming paper.

\subsection{Reducing background}
In section~\ref{sec:nonGaussian_statistic}, we discussed template-dependent reweighting of triggers under the noise hypothesis. Note that reweighting the background differently for triggers from different $(2,2)$ templates was already introduced by Ref.~\cite{CoherentScore}. The difference of our approach is that we do not assume a parametric model for the background distribution (e.g., an exponential curve). Instead, we directly use a smoothed version of the trigger histogram. Another promising avenue to more optimally reweight the background, especially in the high mass case, is using machine learning techniques (see e.g., \cite{Sha22}). We leave exploring this approach to an upcoming work. Note that in our ranking statistic, we do not include information from signal-consistency checks (we currently use signal-consistency checks only to veto triggers \cite{ias_pipeline_o1_catalog_new_search_prd2019}). We also leave incorporating the signal-consistency calculations in our ranking statistic to a future work. 

Apart from the techniques discussed till now, for further reducing the background, we also additionally correct for non-stationarity of PSD and veto noisy triggers. 
For these, we follow methodology similar to that developed for our $(2, 2)$-only searches.
For the vetoes, we currently first subtract our best-fit $(3, 3)$ and $(4, 4)$ waveforms from data and then run the same $(2, 2)$ signal consistency tests as earlier on the residual data (see section~F of \cite{ias_pipeline_o1_catalog_new_search_prd2019}). This is of course an approximation and we aim to modify our procedure to also separately check the consistency of the $(3, 3)$ and $(4, 4)$ parts of the signals in the future.

To account for the non-stationarity of the PSD we use a similar correction factor as in our (2,2)-only search (see \cite{psd_drift}). We first convolve a reference (2,2)-only template from our banks with the data and register its SNR $\rho (t)$. 
We then average the power $\langle \rho^2 \rangle$ within rolling windows of length $\sim 15$\,s and use this factor to account for the variation in the PSD as a function of time in a particular bank.
In the limit when the fluctuations of the PSD over all frequencies are described by a single scalar, this procedure returns an unbiased estimate of the variation of the PSD. Note that we have not currently included (3,3) or (4,4) waveforms in estimating the PSD correction and leave this direction to future work.


\section{Conclusions}
\label{sec:Conclusions}

Nearly all templated gravitational wave (GW) searches only include the quasi-circular quadrupole mode in their templates. We showed that a mode-by-mode filtering method can be used to efficiently and cheaply introduce higher-order harmonics in a templated search. In this method, we first create templates separately for different harmonics (see section~\ref{sec:TemplateBanks} and our companion paper~\cite{Wad23_TemplateBanks}). We then filter each harmonic separately with the data and combine the resulting timeseries using the trigger-ranking statistics derived in section ~\ref{sec:NewDetStats} (see Fig.~\ref{fig:Triggering_modes}). Compared to the brute force method of making templates which have a combination of harmonics, our method has a two-fold advantage: $(i)$ the computational cost of our method is relatively much lower (only a factor of 3 times that of the (2,2)-only search). ($ii$) our method enables efficient marginalization over inclination and initial orbital phase of the binary, and penalizes unphysical degrees of freedom, thereby preventing a degradation in the overall sensitivity of the search (see section~\ref{sec:CoherentScore} and Fig.~\ref{fig:Mode_ratios}). We also include a correction corresponding to non-Gaussian noise to our ranking-statistic (see section~\ref{sec:nonGaussian_statistic} and Fig.~\ref{fig:RankFunction}). 

We run our pipeline on the O3 LIGO-Virgo data and report the detections of new candidate events in our companion paper~\cite{Wad23_HM_Events}. We plan to quantify gain/loss in sensitive volume-time upon adding higher modes in different regions of the parameter space in an upcoming injection study.

The mode-by-mode filtering method and ranking-statistics that we develop could also be useful for including additional degrees of freedom in templates in other scenarios (e.g., to include orbital eccentricity or precession in templates).
The only requirement is that the waveform should be linearly decomposable into the aligned-spin quasi-circular (2,2) mode and other components. While this decomposition is more straightforward for the cases of higher harmonics and eccentricity, it is worth mentioning that recent papers have also explored a similar decomposition for precession \cite{Fai20, McI23, Fai23_SimplePE}. We plan to explore searches including eccentricity and/or precession in future studies. Specific code modules associated with computing the HM marginalized statistics, and implementing the band eraser are available at \url{https://github.com/JayWadekar/GW_higher_harmonics_search/tree/main/Pipeline_modules_arXiv_2405.1740}. Codes and tutorials for running the \texttt{IAS-HM} search pipeline on raw GW strain data are available at \url{https://github.com/JayWadekar/gwIAS-HM}.

\acknowledgments

We thank Mukesh Kumar Singh, Koustav Chandra, Mark Cheung, Bangalore Sathyaprakash and Siddharth Soni for helpful discussions.
DW gratefully acknowledges support from the National Science Foundation and the Keck foundation. 
TV acknowledges support from NSF grants 2012086 and 2309360, the Alfred P. Sloan Foundation through grant number FG-2023-20470, the BSF through award number 2022136, and the Hellman Family Faculty Fellowship.
JR acknowldeges support from the Sherman Fairchild Foundation.
BZ is supported by the Israel Science Foundation, NSF-BSF and by a research grant from the Willner Family Leadership Institute for the Weizmann Institute of Science. 
MZ is supported by NSF 2209991 and NSF-BSF 2207583. 
This research was also supported in part by the National Science Foundation under Grant No. NSF PHY-1748958. We also thank ICTS-TIFR for their hospitality during the completion of a part of this work. 

This research has made use of data, software and/or web tools obtained from the Gravitational Wave Open Science Center (\url{https://www.gw-openscience.org/}), a service of LIGO Laboratory, the LIGO Scientific Collaboration and the Virgo Collaboration. LIGO Laboratory and Advanced LIGO are funded by the United States National Science Foundation (NSF) as well as the Science and Technology Facilities Council (STFC) of the United Kingdom, the Max-Planck-Society (MPS), and the State of Niedersachsen/Germany for support of the construction of Advanced LIGO and construction and operation of the GEO600 detector. Additional support for Advanced LIGO was provided by the Australian Research Council. Virgo is funded, through the European Gravitational Observatory (EGO), by the French Centre National de Recherche Scientifique (CNRS), the Italian Istituto Nazionale di Fisica Nucleare (INFN) and the Dutch Nikhef, with contributions by institutions from Belgium, Germany, Greece, Hungary, Ireland, Japan, Monaco, Poland, Portugal, Spain.

\appendix

\begin{figure*}
\centering
\includegraphics[width= .9\textwidth,keepaspectratio=true]{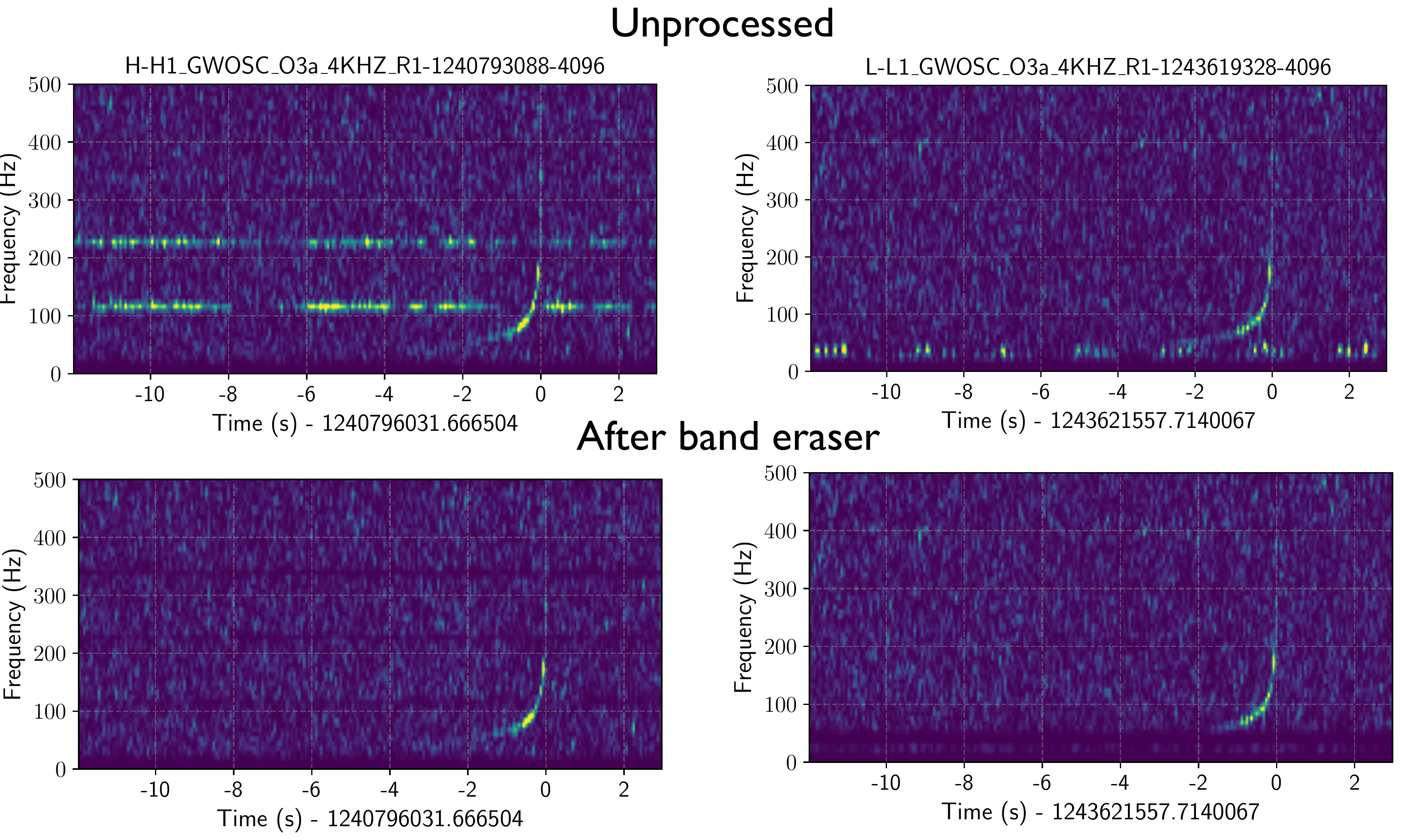}
\caption{In the noisy data file shown in the top panel of Fig.~\ref{fig:BE_specgram}, we inject a low-mass merger waveform ($[m_1,m_2,\chi_{1z}, \chi_{2z}] = [5.23,  4.32, -0.21 ,  0.08]$) and show the spectrogram close to the merger in the top panels. We then apply the band eraser on these files and show the cleaned spectrogram in the bottom panels. We see that most of the power in the signal is recovered while the noisy bands are selectively removed. For reference, we only lose $\sim$ 2\% (9\%) SNR of the injection in the left (right) panel due to overlap with the noisy bands and the rest is recovered.}
\label{fig:BE_injection}
\end{figure*}

\section{Trigger-collection statistics with HM}
\label{sec:single_det_statistic}

In section~\ref{sec:CoherentScore}, we derived the marginalized statistic for ranking multi-detector triggers. In our pipeline, we first collect triggers from individual detectors. We then form a group of triggers with their counterparts in other detectors (we discard triggers which do not have a counterpart). In this Appendix, we derive a single-detector statistic which is quick to compute and allows for triggers with physical combinations of HM. We then generalize our result to quickly identify which multi-detector trigger groups to select for further analysis.

\subsection{Motivation}
One option to collect single-detector triggers is to naively use the incoherent squared SNR: $|\rho^\perp_{22}|^2+|\rho^\perp_{33}|^2+|\rho^\perp_{44}|^2>\rho^2_\mathrm{threshold}$ and leave the work of up-weighting triggers with physical combination of HM to our ranking statistic in Eq.~\eqref{eq:MultiDetCS}. While this would not lead to a loss in the search sensitivity, it will impose a heavy burden on the coincidence section of our pipeline and also increase our memory requirements manyfold.

 The reason for this can be inferred from Fig.~\ref{fig:Mode_ratios}, where we see that naively collecting triggers using $|\rho^\perp_{22}|^2+|\rho^\perp_{33}|^2+|\rho^\perp_{44}|^2$ but using the same value of $\rho_\mathrm{threshold}$ as the $(2, 2)$-only case, the number of collected triggers increases by a factor of over $100$.
The value of this squared-sum statistic in the absence of a signal will be the sum of 6 Gaussian random variables following a $\chi^2$-distribution with 6 degrees of freedom (corresponding to the real and imaginary parts of each modes). Increasing the number of degrees of freedom quickly increases the number of triggers at a fixed collection threshold, e.g., under Gaussian noise, $P(\chi^2_\mathrm{6\, d.o.f}>30)/P(\chi^2_\mathrm{2\, d.o.f}>30) \sim 128$. This can lead to a significant increase in both the time and memory cost of the search. One alternative is to use a higher threshold for $\rho^2_\mathrm{threshold}$ in the case of HM, but this can lead to losing out less-bright $(2, 2)$-dominated physical signals below this threshold.

\subsection{Derivation}
To collect individual detector triggers, let us start with the following approximate version of the multi-detector statistic given in Eq.~\eqref{eq:CoherentScore},
\be\begin{split}
e^{\frac{1}{2}\rho^2_\text{single-det}} =& P(\alpha|\mathcal{S}) \int d\Pi(R_{33},R_{44}, \iota, \phi_0, D, \psi | \alpha)\\
& \times \exp \left[ N_\mathrm{det}\left( \mathrm{Re}\{\langle d | h \rangle\} - \frac{1}{2}\langle h | h \rangle \right) \right]
\end{split}
\label{eq:rho2_SingleDet}\ee
where $N_\mathrm{det}$ will the effective number of detectors later used in the coincidence step of our analysis. $\Pi$ is again the prior over extrinsic parameters and HM mode ratios.

We need the single-detector statistic to be quick to compute as it will be applied to a large number of low SNR Gaussian noise triggers.
To speed up our computation, we maximize (instead of marginalizing) over the extrinsic paramters $\phi_0, D, \psi, \hat{\bm{n}}$, but we retain the marginalization over the amplitudes of HM relative to (2,2) (i.e. over $\iota$ and $R_{\ell m}$). The job of the statistic derived in this section is to downweight triggers with unphysical mode ratios at the single detector stage.

We also use the orthogonalized templates $\h^\perp_{\ell\ell}$ in this subsection to simplify our calculation (see Eq.~\eqref{eq:SNR_timeseries} for their definition).
The expected ratios of SNRs for the orthogonalized modes can be pre-calculated using \texttt{IMRPhenomXHM} waveforms (similar to the case of un-orthogonalized waveforms in Eq.~\eqref{eq:Rlm}) as
\be\begin{split}
&r_{33} \equiv \frac{|\langle h^\perp_\mathrm{33,XHM} | h^\perp_\mathrm{33,XHM} \rangle|}{|\langle h^\perp_\mathrm{22,XHM} | h^\perp_\mathrm{22,XHM} \rangle|}\\
&r_{44}\equiv \frac{|\langle h^\perp_\mathrm{44,XHM} | h^\perp_\mathrm{44,XHM} \rangle|}{|\langle h^\perp_\mathrm{22,XHM} | h^\perp_\mathrm{22,XHM} \rangle|}\\
&w \propto \langle h_\mathrm{XHM} | h_\mathrm{XHM} \rangle^{3/2}
\end{split}
\label{eq:Rlm_perp}\ee

We calculate and store $\{r_{33}, r_{44}, w\}$ samples for each subbank separately. We sample over waveforms $h_\mathrm{XHM}$ corresponding to the intrinsic parameter space associated with each subbank. Note that we also sample uniformly over $\iota$ in $h_\mathrm{XHM}$ in Eq.~\ref{eq:Rlm_perp} whereas, in Eq.~\eqref{eq:Rlm}, we had fixed $\iota=\pi/2$ (we now intend to calculate the integral over both $\iota$ and $R_{\ell m}$ at the same time using the $r_{ii}$ samples).
$w^{(i)}$ is the weight of each sample corresponding to its observable distance (e.g., we give a larger weight to a low inclination system with large $M$ and symmetric $q$ as it is observable to a larger distance, and vice versa). We will use these samples below in Eq.~\eqref{eq:SingleDetScore} for marginalizing over the $r_{33}, r_{44}$ parameter space. Note that $r_{33}, r_{44}$ are functions of $R_{33}, R_{44}, \sin \iota$ (under the approximation that $R_{33} \langle\h_{22}|\h_{33}\rangle \ll 1$ and $R_{44} \langle\h_{22}|\h_{44}\rangle \ll 1$, we recover $r_{33}\simeq R_{33} \sin \iota$ and $r_{44}\simeq R_{44} \sin^2 \iota$).
Upon rewriting the likelihood terms from Eqs.~\eqref{eq:hh} and~\eqref{eq:dh} for the case of orthogonalized modes, we get the RHS of Eq.~\eqref{eq:rho2_SingleDet} to be
\begin{widetext}
\onecolumngrid
\be\begin{split}
\int d\Pi(r_{33},r_{44}, \phi_0, D, \psi | \alpha) \exp \bigg(N_\mathrm{det} \bigg[ \mathrm{Re}\Big\{\mathcal{F}^*\, (\rho_{22}^\perp + e^{i\phi_0}r_{33}\, \rho_{33}^\perp + e^{2i\phi_0}r_{44}\, \rho_{44}^\perp)\Big\} - \frac{1}{2}|\mathcal{F}|^2 (1+r_{33}^2 +r_{44}^2) \bigg]\bigg)
\end{split}\ee
where we used the notation $\mathcal{F}^*\equiv \frac{D_0}{D}\left[ F_{+} \frac{1 + \cos^2 \iota}{2} \, - \, i\cos \iota\, F_{\times}\right] e^{2i\phi_0}$. 
Upon maximizing with respect to the complex number $\mathcal{F}^*$, we get 
$\mathcal{F}_\mathrm{max} = (\rho_{22}^\perp + e^{i\phi_0}r_{33} \rho_{33}^\perp + e^{2i\phi_0} r_{44}\rho_{44}^\perp)/(1+r^2_{33}+r^2_{44})$. Note that this is roughly equivalent to maximizing over $D, \hat{\bm{n}}$ and $\psi$. Substituting $\mathcal{F}_\mathrm{max}$ in Eq.~\eqref{eq:rho2_SingleDet} gives
\be\begin{split}
e^{\frac{1}{2}\rho^2_\text{single-det}} \simeq P(\alpha|\mathcal{S}) \int d\Pi(r_{33},r_{44}, \phi_0 | \alpha) \exp \bigg[  \frac{N_\mathrm{det}}{2 (1+r_{33}^2 +r_{44}^2)} \Big|\rho_{22}^\perp + e^{i\phi_0}r_{33}\, \rho_{33}^\perp + e^{2i\phi_0}r_{44}\, \rho_{44}^\perp\Big|^2 \bigg]
\end{split}\ee
The next step is to explicitly maximize over $\phi_0$ (earlier, $\phi_0$ was degenerate with the phase of the response factor of the detector and only their combination was maximized). In a general case, this maximization depends on the relative amplitudes of the different modes. Note that one could also marginalize of $\phi_0$ using trapezoid quadrature in a similar manner to Ref.~\cite{Rou23_CoherentScore} but we leave this to future work. We approximately do the $\phi_0$ maximization in two regimes. First when the (4,4) mode is weaker than (2,2) and (3,3), i.e., $r_{44} |\rho_{44}^\perp| \ll |\rho_{22}^\perp|$ and $r_{44} |\rho_{44}^\perp| \ll r_{33} |\rho_{33}^\perp|$, in which case $\phi_0^\mathrm{max} = - ( \arg \rho_{33}^\perp - \arg \rho_{22}^\perp)$; similarly for the second case when $(3,3)$ mode is weaker than $(2,2)$ and $(4,4)$. The first case is encountered more often and the corresponding single detector score upon substituting $\phi_0^\mathrm{max}$ becomes

\be\begin{split}
e^{\frac{1}{2}\rho^2_\mathrm{single-det}} &\simeq P(\alpha|\mathcal{S})  \sum_{i}
w^{(i)} \exp \bigg[  \frac{N_\mathrm{det}}{2 (1+[r^{(i)}_{33}]^2 +[r^{(i)}_{44}]^2)} \bigg||\rho_{22}^\perp| + r^{(i)}_{33}\, |\rho_{33}^\perp| + r^{(i)}_{44}\, |\rho_{44}^\perp| e^{i [\arg (\rho_{22}^\perp)+\arg (\rho_{44}^\perp)-2\arg (\rho_{33}^\perp) ]}\bigg|^2 \bigg],
\label{eq:SingleDetScore}
\end{split}\ee
where we perform the last remaining integral over mode amplitude ratios using Monte Carlo integration over the samples we generated.
Therefore, we can use the time-series data of $\rho_{ii}$ to construct an approximate score which is very quick to compute. We provide an example of the marginalized score calculation in a jupyter notebook \nblink.

We first apply a cut on $\rho_\text{single-det}$ from Eq.~\eqref{eq:SingleDetScore} to collect triggers from a single detector. Note that we collect triggers from different templates in the same time window. The next step is to identify triggers among the collected ones that share the same template in multiple detectors and have a time-lag difference that is less than the light crossing time (see section~G of \cite{ias_pipeline_o1_catalog_new_search_prd2019}). Typically, we encounter multiple such triggers in the same time window. We therefore need a criterion to pick the best multi-detector trigger among these for further analysis. One option is to use the coherent score given in Eq.~\ref{eq:MultiDetCS}. But this can be expensive to compute for a large number of multi-detector triggers, thus we use an approximate score by generalizing Eq.~\ref{eq:SingleDetScore} for the multi-detector case as

\be\begin{split}
e^{\frac{1}{2}\rho^2_\mathrm{multi-det}} &\simeq P(\alpha|\mathcal{S})  \sum_{i}
w^{(i)} \exp \bigg[ \\
& \frac{1}{2 (1+[r^{(i)}_{33}]^2 +[r^{(i)}_{44}]^2)} \sum_{k \in \mathrm{detectors}} \bigg||\rho_{22,k}^\perp| + r^{(i)}_{33}\, |\rho_{33,k}^\perp| + r^{(i)}_{44}\, |\rho_{44,k}^\perp| e^{i [\arg (\rho_{22,k}^\perp)+\arg (\rho_{44,k}^\perp)-2\arg (\rho_{33,k}^\perp) ]}\bigg|^2 \bigg]
\label{eq:MultiDetScore}
\end{split}\ee

After this step, we are left with only a single multi-detector trigger in a 0.1 s bucket (we also follow the same procedure on triggers from timeslides). We then use our ranking-statistic in Eq.~\ref{eq:full_det_statistic} to rank the collected multi-detectors triggers and calculate their false alarm rates.

\twocolumngrid
\end{widetext}

\bibliographystyle{apsrev4-1-etal}
\bibliography{HM}
\end{document}